# Single-side access, isotropic resolution and multispectral 3D photoacoustic imaging with rotate-translate scanning of ultrasonic detector array

**Jérôme Gateau,[a,b] Marc Gesnik,[a] Jean-Marie Chassot,[a] Emmanuel Bossy[a]**

a ESPCI ParisTech, PSL Research University, CNRS UMR 7587, INSERM U979, Institut Langevin, 1 rue Jussieu, F-75005, Paris, France

b Laboratoire Kastler Brossel, Université Pierre et Marie Curie, Ecole Normale Supérieure, Collège de France, CNRS UMR 8552, 4 Place Jussieu, 75252 Paris Cedex 05, France

**Abstract (200 words)**. Photoacoustic imaging can achieve high-resolution three-dimensional visualization of optical absorbers at penetration depths ~ 1 cm in biological tissues by detecting optically-induced high ultrasound frequencies. Tomographic acquisition with ultrasound linear arrays offers an easy implementation of single-side access, parallelized and high-frequency detection, but usually comes with an image quality impaired by the directionality of the detectors. Indeed, a simple translation of the array perpendicularly to its median imaging plane is often used, but results both in a poor resolution in the translation direction and in strong limited view artifacts. To improve the spatial resolution and the visibility of complex structures while keeping a planar detection geometry, we introduce, in this paper, a novel rotate-translate scanning scheme, and investigate the performance of a scanner implemented at 15 MHz center frequency. The developed system achieved a quasi-isotropic uniform 3D resolution of ~170 μm over a cubic volume of side length 8.5 mm, i.e. an improvement in the resolution in the translation direction by almost one order of magnitude. Dual wavelength imaging was also demonstrated with ultrafast wavelength shifting. The validity of our approach was shown *in vitro*. We discuss the ability to enable *in vivo* imaging for preclinical and clinical studies.

**Keywords**: photoacoustics, ultrasound array, tomography, multispectral.

**Address all correspondence to:** Emmanuel Bossy, ESPCI ParisTech, PSL Research University, CNRS UMR 7587, INSERM U979, Institut Langevin, 1 rue Jussieu, F-75005, Paris, France; Tel: +33 1 80 96 30 81; Fax: +33 1 80 96 33 55; E-mail: emmanuel.bossy@espci.fr

## 1. Introduction

Photoacoustic (or optoacoustic) imaging has demonstrated high-resolution mapping of optical absorption in biological tissues at depths beyond the operational limit of optical microscopy (1-3). This multi-wave imaging modality relies on the detection of broadband ultrasonic waves generated through thermo-elastic stress following the absorption of transient illumination. Tomographic ultrasonic detection and image reconstruction methods enable the visualization of tissue chromophores and exogenous optical absorbers with a spatial resolution associated with medical ultrasound frequencies (4). The spectral specificity of optical absorption can be exploited to discriminate various absorbers through the use of multiple excitation wavelengths (5). Endogenous absorbers such as hemoglobin enable morphological and functional imaging, in particular blood vasculature (6-8) and local oxygen saturation (9, 10). The bio-distribution and follow-up of exogenous absorbers can reveal additional anatomical and functional information such as blood perfusion (11, 12), drainage in lymph vessels (13), and accumulation of contrast agents (10, 14, 15). Much recent research has been directed towards preclinical studies in small animals (16, 17), but photoacoustic imaging has also promising clinical applications, such as diagnostic of breast cancer (18) and dermatology (19, 20).

Photoacoustic (PA) tomography is highly scalable (3). The upper cutoff frequency limit of the ultrasonic detectors sets the diffraction-limited resolution (21) and the attenuation-limited penetration depth. Typically, central frequencies of 3-8 MHz have showed capabilities of imaging whole-body small animal (16) and human (7) at a few hundreds of micrometers resolution and penetration depths of several centimeters. Detectors with center frequencies > 10

MHz can achieve penetration depths limited to approximately 1 centimeter, but their sensitivity to small features and their higher resolution have recently raised a great interest for photoacoustic imaging at mesoscopic scale (1, 13, 14, 22-25).

For preclinical and clinical biomedical applications at mesoscopic scale, a photoacoustic imaging system should ideally provide spatial performance such as: single-side access, large field-of-view, uniform three-dimensional resolution on the order of the smallest detected ultrasonic wavelength, and visibility of structures larger than the resolution limit regardless of their orientations. Three dimensional (3D) imaging would be preferred because of the inherent volumetric illumination and ultrasound detection, and the more accurate comparison of images acquired at different time points in longitudinal studies. Single-side access allows PA imaging regardless of the topology of the surface and the size of the sample, but limits the achievable angular aperture. A partial angular coverage results in degraded resolution and limited view effects (26). The angular aperture can also be further limited by the spatio-temporal properties of the detectors (27). Typically high sensitivity in ultrasonic detectors comes with directionality and a large area which prevents parallelized detection with the dense spatial sampling required for high frequency PA imaging. Synthetic aperture approaches can however compensate for the directivity (28) and limited parallelization performance of the ultrasonic detection to the detriment of the acquisition time. Consequently, for single-side access PA imaging at mesoscopic scale, the imaging performance results from a compromise between spatial quality and frame rate. Different technical approaches have already been proposed for the high-frequency ultrasonic detection, but lead to quite unbalanced compromises.

Recently, Dean-Ben et al (29) have shown real-time multispectral 3D PA imaging *in vivo* on humans using parallel acquisition on a 256-elements hemispherical array of directional detectors. While attractive for a center frequency of 4 MHz, scaling the approach for high frequency detection will require intensive technical development to build a dedicated detector array and will result in a limited field-of-view because of the natural focus of the detection geometry. Using a planar geometry and an optical method providing broadband (up to 20MHz) and isotropic ultrasound detection (25), Zhang and his colleagues have shown high-resolution single-wavelength images of tumor vasculature in mouse models (6) and microvasculature of human palm (30). However, the employed detection method required scanning point-by-point the detection surface, which results in lengthy acquisitions even with a high repetition rate laser. The same issue occurs for raster-scanned single-element detectors (22, 31). For parallel acquisition of the PA signals along one dimension of the planar detection surface, several groups have implemented systems using ultrasound linear array. High frequency ultrasound linear arrays have been developed for ultrasound imaging and are widely available commercially for center frequency up to 50 MHz (32). Different scanning geometries of linear arrays have been implemented for single-side access 3D PA imaging. First, a single translation of the high-frequency array perpendicularly to its length axis has been used to obtain stacks of 2D images, - each 2D images corresponding to a position of the array (13, 24)-, or to reconstruct 3D images with a better contrast using a synthetic aperture approach (14). Although enabling fast acquisitions and a large field-of-view especially along the direction of the scan, this simple scanning geometry achieves a poor resolution along the translation direction and a very limited view due to the small angular aperture of the linear arrays in this direction (24). To improve the resolution, Schwarz et al (33) proposed to use a bidirectional translation scan (with a 90° rotation around the bisector of the array in the median plane) combined with a multiplicative image processing method. While the method demonstrated an isotropic transverse resolution for absorbing spheres, the multiplicative processing severely reduces the visibility of directive structures to structures detectable for both the scanned directions; i.e. emitting in the intersection of the angular apertures. Conversely, Gateau and colleagues (28, 34, 35) have shown that combining a translate-rotate scanning geometry (with a rotation axis parallel to the length axis of the array) and 3D reconstructions with all the signals can both extend the angular aperture of ultrasound linear arrays and improve the resolution. However, their implementations of the translate-rotate geometry required enclosure of the imaged sample, thereby limiting accessibility to small body parts. In parallel, Preisser et al (36) performed translation scans of an ultrasound linear array under different inclination angles to showcase the limited view and the effect of the inclination. While a 7.5 MHz center frequency (< 10 MHz) linear array was used and only stacks of 2D images with sparse inclination angles were obtained, the imaging set-up presented a single-side access configuration enabling combined translation and rotation of a linear array. We propose herein to implement a resembling single-side access configuration to perform 3D photoacoustic imaging at mesoscopic scale using a high frequency ultrasound array in planar detection geometry.

We investigate in this paper the operation of a novel single-side access rotate-translate PA scanner, benefiting from both parallelized acquisition over the elements of a conventional linear array, and the extension of angular aperture over a large field-of-view provided by advanced combination of translations and rotations (34). The system is based on a 15 MHz ultrasound array, mounted on high-precision motorized stages. In addition to parallelized acquisition, time efficient acquisition using continuous data acquisition (34, 37) are demonstrated while satisfying spatial sampling requirements. The system was characterized *in vitro* with phantoms and the performance compared to translation-only scanner. Moreover, we investigated the multispectral imaging capabilities of our approach with a fast wavelength switching method, as low time difference between the laser pulses at different wavelengths has proven beneficial for reduction of motion artefacts (38, 39).

## 2. Materials and Methods

The PA imaging system presented herein was developed to accommodate samples of all sizes, and therefore achieves ultrasound detection with a single-side access. For the synthetic aperture acquisitions, the detector motion, provided by the rotate-translate scanning system, is constrained to a plane, and the sample is placed in the vicinity of the detection plane.

*2.1. Experimental set-up*

The experimental set-up used in this study is presented in Fig. 1. The PA imaging system consists of four main components: 1) the illumination part comprising a nanosecond-pulsed laser and a fiber bundle, 2) the multi-element ultrasound detector and its data acquisition system, 3) the scanning system with motorized rotation and translation stages, and 4) the acoustic coupling part including a watertight array holder sealed on the soft bottom of a water tank.

A tunable (680–900 nm) optical parametric oscillator laser (Laser SpitLight 600 OPO Innolas, Germany) delivering < 8 ns duration pulses with a pulse repetition frequency (PRF) of 20 Hz and a fiber bundle with two arms (CeramOptecGmbH, Bonn, Germany) were used to generate and guide the excitation light to the sample. Acquisitions performed with a single wavelength were done at 700 nm, and multispectral acquisitions used two wavelengths: 680 nm and 800 nm. The laser technology enabled ultrafast wavelength shifting, which means that the wavelength of the illumination could be shifted to arbitrary values between two consecutive pulses according to a programmed sequence. This feature was used during dual-wavelength acquisitions to alternate from pulse-to-pulse between the two wavelengths. The total per-pulse light energy was measured at the output of the fiber buddle to be 2.5 mJ at 680 nm, 7.5 mJ at 700 nm and 15 mJ at 800 nm.

Illumination was performed from two sides of the sample (Fig. 1 (a)), distributing light over several centimeter squares. The light fluence was therefore below the maximal permissible exposure (40). To avoid engineering waterproof cable entries in the water tank and because of the limited flexibility of the optical fibers, the illumination was performed, in the current experimental set-up, from the upper part of the phantom. For each laser pulse, the intensity fluctuations were monitored with a pyroelectric device incorporated inside the laser, and were used to normalize the measured PA signals. The illumination was kept fixed during the acquisitions and for all the samples.

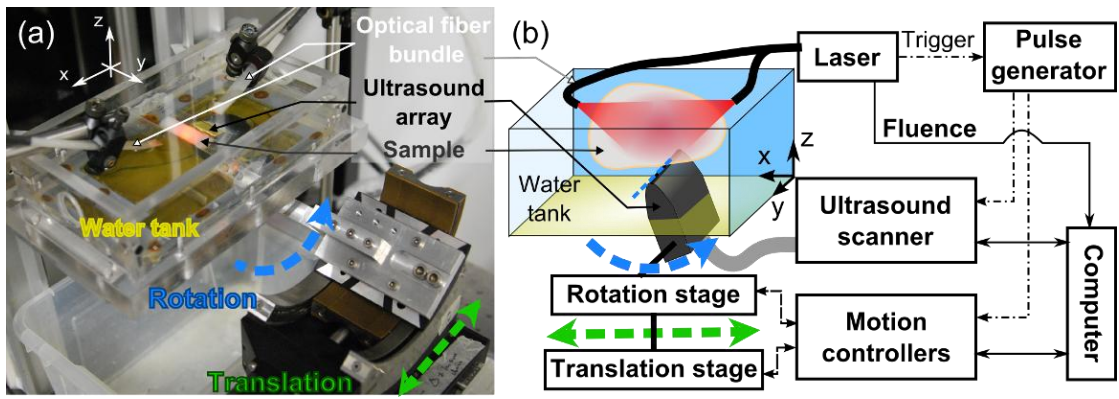

Fig. 1 Experimental set-up. (a) Annotated picture of the PA rotate-translate scanner. The ultrasound array is mounted on the rotation stage and the resulting assembly is translated by the translation stage. The sample is placed above the array of detectors. This configuration is made possible by the use of a water tank with a soft bottom. (b) Schematic drawing of the experimental set-up. A Cartesian coordinate system fixed to the water tank is specified. The x-axis corresponds to the axis of the translation stage. The origin of the system (not shown here) corresponds to the center of the array when the rotation stage and the translation stage are at their middle position, respectively.

Ultrasound detection of the PA waves was performed with a 128-element linear array (Vermon, Tours, France) driven by a programmable ultrasound machine used in a receive-only mode (Aixplorer, Supersonic Imagine, Aix-en-Provence, France). The elements of the array have an averaged center frequency of 15 MHz and an averaged two-way (pulse echo -6 dB) bandwidth of 50%. The elements have a height of 1.5 mm and are cylindrically focused at a focal distance of 8 mm. The pitch of the array is 101 µm. Signals were acquired simultaneously on 128 channels and digitized at a sampling rate of 60 MS/s.

Two high precision motorized stages were used to move the array across the region of interest in a rotate-translate mode: a translation stage (M-605.2DD, Physik Instrumente, Karlsruhe, Germany) and a goniometer stage (WT-90, PI Micos, Eschbach, Germany). The maximum travel range of the stages was 50 mm and 90° for the translation and rotation stages, respectively. Each stage was equipped with an integrated motion encoder, which provided readout of the motor positions during continuous motion. The bi-directional repeatability of the stages was ± 0.2 µm and ± 0.02° for the translation and rotation stages, respectively. The motion controllers (C-863 Mercury Servo Controller Mercury, Physik Instrumente, Karlsruhe, Germany) were customized to record the motor positions at the time of arrival of an external trigger. The data buffer could store the positions corresponding to 1024 successive external trigger signals in an internal memory. This triggered readout of the motor positions enabled a reliable assessment of the positions of the detector array during tomographic measurements of the high-frequency ultrasound field, which is crucial for the image reconstruction. The stages were assembled so that the goniometer stage was translated by the translation stage. The detector array was mounted on the goniometer stage with a rigid arm and a custom-build holder. The length axis of the linear array was mechanically aligned to match the rotation axis of the goniometer. This alignment ensured that the motion of the centers of the elements is confined to a xy-plane.

Acoustic coupling between the sample and the detector array was obtained via a water bath. In this study, the samples were fully immersed to ensure acoustic coupling in all the directions probed by the detector, but for large samples only the surface around the volume of interest could be in contact with the water. Since the active surface of the detector point upwards, the bottom of the water tank must enable both water confinement and motion of the detector inside water. To handle these two constrains, the water tank was equipped with a soft bottom made of a 0.33 mm thick latex sheet (Supatex Heanor, United Kingdom), and a watertight array holder, made of ABS, was sealed on this soft bottom. The array holder was inserted in the perforated latex sheet, and the edges of the hole were then compressed on the outer contour of the holder with knife-edge sealing. The other tank walls were rigid and made of plexiglass.

### 2.2. Automation of the acquisition

The automation of the acquisition was made possible by the stand-alone functionality of the motion controllers and the ultrasound machine. The motion of the motorized stages was macro-programmed in the controllers, and a trigger signal on one of the input ports started the motion sequence. For the ultrasound machine, the imaging sequence was programmed to enable memory allocation, and the signal acquisition was externally triggered to ensure synchronization with the laser emission.

The laser was run continuously and triggered a digital delay and pulse generator (BNC Model 565, Berkeley Nucleonics Corporation, San Rafael, CA) at each laser pulse. When manually enabled, the pulse generator duplicated the trigger signal on three outputs. One output triggered the data acquisition on the ultrasound machine and the two others started the programmed motion of the motorized stages and triggered the readout and storage of the motor position, respectively. Only 1024 successive triggers were duplicated before the digital delay-and-pulse generator was disabled, in order to match the buffer size of the motion controller. Data from the ultrasound machine and the motion controller was then collected on a computer for image reconstruction.

*2.3. Continuous acquisition*

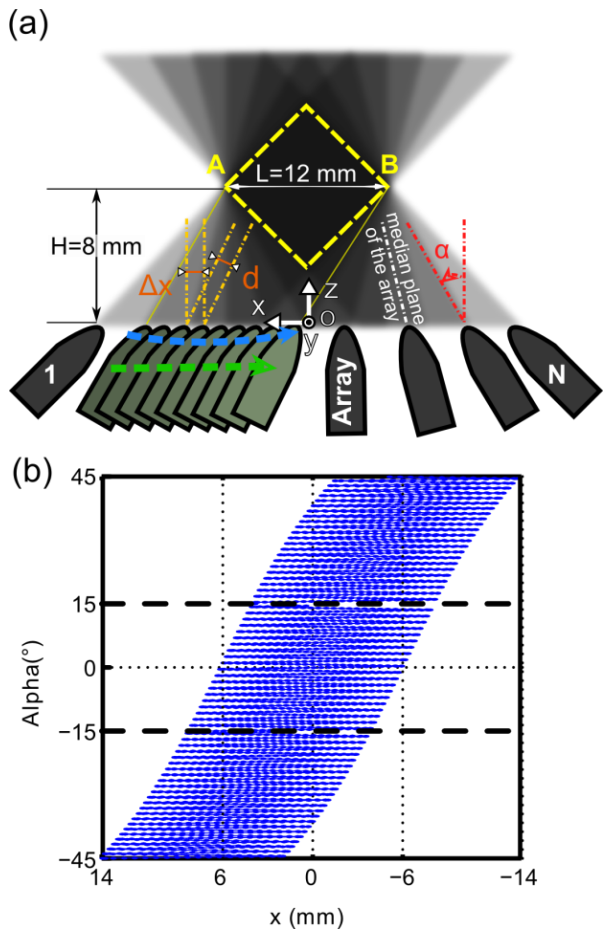

Fig. 2 Continuous scanning parameters. (a) Schematic drawing of the scanning process. Several positions of the ultrasound array are shown including the ones corresponding to the first and the last (N) laser pulses considered. Positions of the array along a one-way translation and corresponding to successive laser pulse events are presented in dark green. The spacing along the x-axis between two successive positions is $\Delta x$. The extreme positions for the one-way translation #i ($x_{i,left}$ and $x_{i,right}$) are shown to be chosen so that the medial plane of the array is confined between the points A and B. The overlap between seven probed volumes (each corresponding to a one-way translation travel) is illustrated in the xz-plane by the superposition of light gray trapezoids. The horizontal sides of these trapezoids correspond to the array translation range for a given one-way translation travel and the other pair of sides illustrates the median imaging plane of the array at the beginning and the end of the travel. The dashed yellow square indicates the area where the overlap between the probed volumes for all the one-way translation of the scan is expected to be maximal. (b) Experimental measurement of the coordinates of the rotation and translation stages for a scan where L=12 mm and a total of N=3072 laser shots were used. To enable the measurement with the motion controllers, the scan was divided into three parts indicated with the dashed lines.

*2.3.1.Scanning geometry*

To acquire tomographic datasets, a rotate-translate scanning motion of the detector array was implemented (Fig. 2). As the elements of the array are cylindrically focused, the sensitivity pattern of the detector is directive and the volume of higher sensitivity is confined around the median plane of the array (27). The translation motion aims at scanning the focal area across the volume of interest along the x-axis. The rotation motion modifies the orientation of the median plane to probe the PA waves along different directions in the xz-plane, and therefore increase the angular aperture along the x-axis (28).

A scanning procedure in which the array is continuously and simultaneously rotated and translated was chosen. This scanning method has been shown to be time-efficient compared to step-by-step scanning as the acquisition speed is then primarily limited by the laser pulse repetition frequency (PRF=20Hz here) (14, 34, 37). The time for communication with the motion controllers or motion between successive steps does not slow down the acquisition. To systematically scan the volume of interest and avoid redundant information (similar orientation and translation positions), we chose to define a slow stage, for which the motion range is travelled only once during the entire scan, and a fast stage, for which the motion range is travelled repetitively in oscillatory motion (Fig. 2 (b)). Because its maximum speed does not allow fast repetitive travel over the 90° rotation range, the goniometer stage was defined as the slow stage. Full datasets were obtained by translating the detector array across the sample repetitively and rotating the detector array from -45° to 45° to obtain the largest possible angular aperture. Since the orientation of the median plane changes constantly, the probed volume is different for each one-way translation travel. The endpoints for each new travel across the sample should be adjusted to set the overlapping subset of all volumes imaged with translation scans of different orientations. For this study, we chose to adjust the endpoints for each one-way translation travel so that the median plane of the array moves along the x-axis within a fixed interval (segment AB on Fig. 2 (a)) located inside the sample at the z-coordinate H = 8 mm. 8 mm corresponds to the focal distance of each element of the array and therefore to the region of highest sensitivity. Consequently, the overlapping subset of the volumes probed for the different translation travels had a square cross section parallel to the xz-plane – the segment AB being a diagonal of the square (Fig. 2 (a)). For every one-way translation, the translation range was fixed and equal to L = 12 mm.

*2.3.2.Spatial sampling*

The geometrical characteristics of the cylindrically focused elements result in a sensitivity field with calculated full width half maximum (FWHM) dimensions in the xz-plane at 15 MHz of ∼ 530 µm around the focus in the transverse dimension, and about 2 cm for the depth-of-field (27) . Therefore, the minimum spatial sampling along the x-axis (translation direction) should be set so that the sensitivity patterns for two successive laser shots are contiguous. Herein, the motion parameters were set so that the distance between the positions of the median plane of the array for two successive laser shots was d = 500 µm. Because of the orientation of the array, the corresponding travel distance along the x-axis is: $\Delta x(\alpha) = d/\cos(\alpha)$ where α is the angle between the median plan and the yz-plane (Fig. 2 (a)). Since α is continuously changing, the translation speed $v = \Delta x(\alpha) \times PRF$ should ideally be constantly modified in a non-linear way. Practically, the translation speed was set constant for each one-way translation travel, and adjusted between successive travels as explained in the next section.

For the rotation stage, the angular spatial sampling was chosen to match the pitch of the array. A pitch of 0.101 mm means that the spatial sampling of the array is about the ultrasound wavelength at 15 MHz. If we convert this spatial sampling in angular sampling, at a distance of 8mm (i.e. the focal distance) from the array, the mean angular sampling is about 0.7° if we consider the 32 closest element of the array. For a rotation range of 90°, achieving such an angular sampling means that each point should be seen by 128 successive angles regularly spaces. For our geometry, this criterion corresponds to a minimum of 128 one-way translation travels.

*2.3.3. Scanning procedures*

Three different scanning modes were employed in this study (Table 1). The first one was a translational scan of the array, without any rotational motion. Therefore, the goniometer stage was kept fixed at α=0. This scanning mode was used to compare the proposed rotate-translate scan with previous single-side access implementations (13, 14, 24). Only single-wavelength acquisitions were performed in this first mode. The translation speed and the translation range for each one-way motion were set for all the translation travels to: $v_0 = d*PRF = 10$ mm.s$^{-1}$ and [-L/2 ; L/2], respectively. Because no time averaging of the signals can be performed due to the continuous motion, multiple one-way translation travels were done to improve the signal-to-noise ratio in comparison to a single one-way travel. The second and third scanning mode were used to perform single-wavelength and multispectral acquisitions with the proposed rotate-translate scanning geometry, respectively. In the third mode, the number of laser shots used to acquire the complete dataset corresponds to the product of the number of wavelengths $n_\lambda$ and the number of laser shots in the second scanning mode. Indeed, the sampling requirement should be verified for all the wavelengths. To reduce the difference that can appear between 3D images due to motion of the sample (38), we chose here to interlace the acquisitions for the different optical wavelengths. For the dual-wavelength acquisition implemented here, the laser was programmed to send on a pulse basis alternatively one of the two different wavelengths, the translation speed was set so that the distance between the positions of the median plane of the array for two successive laser shots was d/2, and the rotation speed was divided by two.

For the practical implementation of the second and third acquisition mode with the motion controllers, several choices we made. First, because the recording capacity of the controllers was limited to 1024 positions, the motion was segmented into n parts (n=3.$n_\lambda$ here, with $n_\lambda$ the number of wavelengths) to obtain the full tomographic acquisition. Second, the rotation speed was chosen constant and the rotation endpoints were set to scan a 90°/n range (Table 1). To macro-program the translation motion in the limited internal memory of the controller memory, the speed of the translation and the translation endpoints were computed considering a fixed angle for each one-way translation travel instead of continuously adjusting the parameters with the actual angle α. The angle was taken equal to the angle at the end of the translation motion. For the one-way translation motion number $i$, the angle is called $α_{i,\,max}$. The acceleration and deceleration of the translation stage were set to their maximum value: 150 mm.s$^{-2}$, and the translation speed and endpoints were calculated by taking into account the pulse repetition rate of the laser and the trapezoidal velocity profile. The successive translation speed and endpoints values were stored in the controller memory and the motion was programmed to change direction, endpoint and speed every time an endpoint was reached.

Table 1 Motion parameters for the different modes with PRF= 20Hz and H= 8 mm. The measured number of one-way translations was not directly proportional to N/ $n_\lambda$ mostly because of the acceleration and deceleration phases and the change in the speed to travel along the translation range L.

| Scanning mode | Single-wavelength translate-only | Single-wavelength rotate-translate | Dual-wavelength rotate-translate |
|---|---|---|---|
| Number of measurement positions (N) | 1024 | 3072 | 6144 |
| Number of wavelengths ($n_\lambda$) | 1 | 1 | 2 |
| Number of parts (n) | 1 | 3 | 6 |
| Rotation range | 0 deg | 90 deg | 90 deg |
| Rotation speed (deg.s$^{-1}$) | - | 90*PRF/N  0.59 deg.s-1 | 0.29 deg.s-1 |
| Rotation endpoints for the part # j (with $j \in [\![1,n]\!]$) | - | [ -45deg +(j-1)/n*90° ; -45deg +j/n*90°] | |
| Translation range (L) | | 12 mm | |
| Translation endpoints for the one-way translation #i | $x_{i,left} = -L/2$  $x_{i, right} = L/2$ | $x_{i,left} = -L/2 + H*\tan(\alpha_{i, max})$  $x_{i, right} = L/2 + H*\tan(\alpha_{i, max})$ | |
| Translation speed for the one-way translation #i | $v_0$= 10 mm.s$^{-1}$ | $v_0/\cos(\alpha_{i, max})$ | $v_0/(2*\cos(\alpha_{i, max}))$ |
| Measured number of one-way translations | 40 | 132 | 138 |
| Total acquisition time : N/PRF | 51.2 s | 2.56 min | 5.12 min |

*2.4. Signal processing and image reconstruction*

Datasets acquired with the PA scanner were arranged in four-dimensional matrices. The matrix dimensions were: the time, the elements of the array, the translation positions, and the rank of one-way translations. A fifth dimension was added for multispectral acquisitions. This dimension corresponds to the different wavelengths of the successive laser shots. The signals were sorted depending on the corresponding wavelength and the four-dimensional matrices for each wavelength were processed independently.

Before reconstructing 3D images with a backprojection algorithm, signals processing was performed along each of the four dimensions. Along the first dimension, signals were bandpass filtered (Butterworth order 3, cutoff frequencies 5 and 20 MHz) for noise removal. Along the three other dimensions, different apodization functions were applied. For the second dimension, the elevation aperture along the length axis of the array (y-axis) was kept constant and equal to 0.6 as long as possible with a dynamic aperture approach, as described in (28). This dynamic aperture aims at keeping the elevation angle of acceptance constant for all the points in the 3D image. A 20% tapered cosine (Tukey) window was used to apodize this dynamic aperture, and to avoid strong side lobes. For the third dimension, because the acceleration/deceleration phases of the translation stage enhance locally the spatial sampling, a window with a 20% cosine taper from 1 to 0.3 was applied to the signals for each one way travel. Such a window prevents the emphasis of signals arising from the better sampled areas. Along the fourth dimension, a window with an 80% cosine taper from 1 to 0.1 enabled apodization over the angular aperture induced by the rotation stage to avoid related side lobes. This last apodization is necessary because of the limited view of the detection geometry. Because there is no rotation in the single-wavelength translate-only acquisition mode, no apodization was performed along the fourth dimension for reconstruction in this mode.

Because of its slow motion (maximum translation speed of 14 $mm.s^{-1}$, and rotation speed of 0.59 $deg.s^{-1}$), the ultrasound array can be considered quasi-static during the PA wave propagation even if the array is continuously moving: with a maximum propagation distance for the ultrasound waves of 24 mm, the maximum displacement of the focal spot of the array during the wave propagation (including translation and rotation) is ~ 0.2 µm and could therefore be considered as negligible.

Images were reconstructed on a grid composed of cubic voxels of side 25 µm. The spatial extent of the image in the three dimensions was: from -6 mm to 6 mm along the x-axis, from 2 mm to 14 mm along the z-axis to comprise the cross-section parallel to the xz-plane of the overlapping subset of the probed volumes (marked with a yellow dashed square on Fig. 2), and from - 4.25 mm to 4.25mm along the y-axis. The image length along the y-axis was chosen to limit artefacts caused by the finite aperture of the array (array length of ~13 mm), and to match the side of the dashed square and therefore to obtain a cubic volume were the angular aperture is expected to be maximal.

A backprojection algorithm in the far field approximation, described in details in (28), was used for image reconstruction. This backprojection algorithm is a heuristic reconstruction method based on the phenomenological description, given in (4), of Xu and Wang's 3D backprojection algorithm (41). For the sake of clarity, the main features are summarized here. First, the first time-derivative of the processed signal: $-\partial V/\partial t$ was backpropagated radially onto the image grid assuming point-like detectors located in the centers of the elements of the ultrasound array. Neither the cylindrical focusing of the array nor the rotation angle was taken into account here. The non-derivative term V/t was found negligible with respect to the derivative term. Then, the reconstruction of the images was performed using all the detection signals. However, to enable the dynamic aperture reconstruction, each plane at constant y-coordinate was reconstructed independently. Although not optimal for focused detectors, this backprojection algorithm was chosen here because it is fast, memory efficient and robust. Moreover, its simple computational steps allow processing of large datasets. Furthermore, this reconstruction algorithm has already been shown to lead to high quality photoacoustic images with imaging systems using rotated and translated linear ultrasound arrays (23, 28, 34). The speed of sound used for the reconstruction was determined experimentally to be: $c = 1490$ m/s.

Because of the electric impulse response of the transducer and the reconstruction algorithm, non-physical negative values appeared in the image. To obtain images with positive values only, we used an envelope-detection approach. First, the Hilbert transform was applied to the acquired time signals to obtain 90 ° phase shifted signals. Then, the acquired signals and the quadrature signals were backprojected independently to form two 3D images. Finally, the envelope-detected 3D image was obtained by computing for each voxel the root mean square of the voxel values in the two images.

The 3D images were visualized using maximum amplitude projection (MAP) images and a rotation around one axis. This visualization was performed with the 3D project option of the software ImageJ (42). To improve the image contrast before visualization, the voxel values were thresholded. Image postprocessing was based on the assumption that the majority of the reconstructed voxels consisted of noise, since the actual absorbing targets occupy only a small part of the total volume in the region of interest. First, a histogram of the image values was computed. Voxels with values below the one corresponding to the maximum of the histogram were discarded as noise. For the remaining voxels, the first 10% to 50% was discarded to limit the background noise from being displayed.

*2.5.Imaging phantoms*

For the characterization of the system, different calibrated absorbers were embedded in turbid gel and molded to form cylinders of 12-mm diameter gel. A turbid gel mimicking the speed of sound in soft tissues was prepared by mixing 1.8% w/m agar gel (Fluka analytical) with 8% v/v intralipid-10%. The unmolded cylinders were inserted in a fixed sample holder so that the embedded samples faced the ultrasound array.

To study the resolution and spatial homogeneity of the images, three phantoms comprised of 50µm-diameter black microspheres (Black Paramagnetic Polyethylene Microspheres 1.2g/cc 45-53um, Cospheric LLC, Santa, Barbara, CA) were prepared. With the laser pulse duration used here, the absorbing microspheres are expected to generate PA signals with a peak frequency greater than the highest recorded frequency (43) and are therefore suited for resolution assessment. For each phantom, the microspheres were randomly spread and trapped over a surface of melted turbid agar gel, and after solidification, the plane of microspheres was embedded in more gel. The spatial distribution of the absorber along planes allows comparing PA images with optical pictures taken before embedding the sample. For each phantom, the orientation of the plane was set to be orthogonal to an axis of the cartesian coordinate system to cover the main orientations of the system. The spheres were mostly located in the xz-plane for Phantom 1, in the xy-plane for Phantom 2 and in the yz-plane for Phantom 3.

To study the capabilities of rotate-translate scanning geometry to image complex shapes, and in particular samples comprised of directive structures with multiple orientations such as a vascular network, a strip of black leaf skeleton arranged in a helical ribbon shape was embedded in turbid gel (Phantom 4). The smallest branches of the leaf vascular network were approximately 50 µm wide.

To demonstrate the multispectral capabilities of the current implementation, four polycarbonate capillary tubes with a 100 µm inner diameter (200 µm outer diameter, CTPC100-200, Paradigm Optics, USA) were inserted in gel (Phantom 5). Two were filled with black indian ink (Sennelier à la pagode, France), and the two others were filled with a water solution of brilliant blue (colorant E133, scrapcooking, France). The black indian ink and the blue dye were diluted to have a similar optical density (OD) of about 5 at 680 nm. The OD of the solutions was measured with a scanning spectrophotometer (Genesys 10S UV-Vis, Thermo Fisher Scientific Inc., WI). At 800 nm, the Brilliant blue solution had an OD of 0.15 (4.8 at 680 nm) and the Black indian ink had an OD of 4.0 (4.9 at 680 nm). Brillant blue was chosen here because it has a similar spectrum as patented blue and methylene blue used in clinical practice to color lymph vessels for instance.

*2.6. Resolution study*

To assess the three-dimensional resolution in Phantom 1, Phantom 2 and Phantom 3, the images of the microspheres were fitted with 3D Gaussian blobs using a non-linear least squares minimization. The three main axes of the 3D Gaussian blob were the axes of the Cartesian coordinate system and the following equation was used:

$$g_{3D\ Gaussian\ blob} = A.\exp\left(-\left(\frac{(x-x_0)^2}{2.\sigma_x^2} + \frac{(y-y_0)^2}{2.\sigma_y^2} + \frac{(z-z_0)^2}{2.\sigma_z^2}\right)\right), \quad (1)$$

where $x_0$, $y_0$ and $z_0$ are the coordinate of the center of the Gaussian blob, A is the amplitude, and $\sigma_x$, $\sigma_y$, $\sigma_z$ are the Gaussian RMS width along the axes x, y and z, respectively.

The parameters were determined using cubes of 11 x 11 x 11 voxels, centered on the images of the spheres. The size of the cube was determined in order to avoid interference of background signals. The full width half maximum (FWHM) resolutions were computed using the equation:

$$R_{FWHM} = 2.\sqrt{2.\ln(2)}.\sigma$$

## 3. Results

*3.1. Resolution and field-of-view with isotropic objects*

Fig. 3 displays the reconstructions of Phantom 1, Phantom 2, and Phantom 3, composed of 50 µm-diameter microspheres, for acquisitions in the single-wavelength rotate-translate scanning mode. The comparison between the optical pictures of the samples (Fig. 3 (A)) and the corresponding PA images (Fig. 3 (B)-(C)) shows the spatial fidelity of the obtained PA images in terms of location of the absorbers. This spatial fidelity is observed in the cubic volume with a cross-section parallel to the xz-plane indicated with a dashed-line square (Fig. 3 (I)). Smearing can be noticed outside of the cubic volume because of partial overlapping of the volumes probed for the different translation travels (Fig. 3 (I-B)). Fig. 3 (C) shows the confinement of the microspheres in planes.

For a quantitative assessment of the homogeneity of the reconstruction inside the cubic volume in terms of both amplitude and resolution, 20 to 25 well-separated microspheres were chosen for each phantom (Fig. 3 (A)), and the image portions around the individual microspheres were fitted with 3D Gaussian blobs (Fig. 3 (D)). The amplitudes of the reconstructed microspheres were found consistent for the different locations within each phantom. Variations in amplitude can be caused by the dispersion in size and absorption of the microspheres, and the spatial heterogeneities of both the fluence and the sensitivity field of the detection. For Phantom 1 and Phantom 3, a dependency of the reconstructed amplitude with the z-coordinate can be noticed. This variation can be attributed to the fluence inhomogeneity for the lower part of the phantoms and to the sensitivity pattern of the array for the upper part. Indeed, the illumination was on the upper part of the phantom (Fig. 1), therefore, microspheres located deeper in the water tank (low z) received less optical energy. It worth noting that the amplitude reduction is not associated with a lower resolution here, and is therefore not expected to be caused by a smearing effect in the reconstruction. For microspheres with z-coordinates above 12 mm, the distance to the array is large for all the scanned positions and the sensitivity lower. Neither the fluence variation nor the spatial response of the detector along the median plane of the array was taken into account in the reconstruction algorithm. Phantom 2 for which the microspheres had about the same z-coordinate has the lowest standard deviation for the amplitude. For the resolution, we found an almost isotropic resolution on the order of 170 μm for the different locations explored over the three phantoms. Along the x-axis, the FWHM resolution $R_{FWHM\ x}$ is consistent over the three phantoms. The resolution in this direction is the highest for each phantom because of the large angular aperture (90°) provided by the rotate-translate motion. We note the low standard deviation of $R_{FWHM\ x}$ for a constant z-coordinate (Phantom 2 Fig. 3 (II-D)) and a degradation of the resolution for microspheres located at z > 10mm (Phantom 1 and Phantom 3). This degradation is most probably due to the lower sensitivity of the focused detector for absorbers far from the focus that reduces the effective angular aperture. Along the y-axis, the FWHM resolution $R_{FWHM\ y}$ is also consistent over the three phantoms, although it is lower for Phantom 3. The resolution in the y-direction is the lowest because the limited aperture given by the length of the array. The angular aperture in this direction is ~80° for a f-number of 0.6 . $R_{FWHM\ y}$ is maintained constant with the growing aperture adjustment but degrades for microspheres with a non-constant y-coordinate and a z-coordinate above 10mm (Fig. 3 (III)) as the length of the array implies a degraded f-number. The standard deviation of $R_{FWHM\ y}$ is similar for the three phantoms. Along the z-axis, the FWHM resolution $R_{FWHM\ z}$ is similar for the three phantoms. The resolution in this direction is mainly related to the electric impulse response of the ultrasound detector and the envelop detection.

In the translation-only scanning mode, the three-dimensional resolution could only be assessed on Phantom 3 because the elongated shape of the reconstructed microspheres along the x-axis did not enable discrimination of individual microsphere in Phantom 1 and Phantom 2. Because of the strong anisotropic resolution, the approximation by a 3D Gaussian blob was no longer valid. Therefore, $R_{FWHM\ y}$ and $R_{FWHM\ z}$ were assessed by fitting the middle cross-section of the imaged microspheres with a 2D Gaussian surface, and $R_{FWHM\ x}$ was assessed on the central 1D profile along the x-axis. We found $R_{FWHM\ x} = 1460 \pm 13$ µm (mean ± standard deviation), $R_{FWHM\ y} = 203 \pm 12$ µm and $R_{FWHM\ z} = 163 \pm 10$ µm. The FWHM resolution along the x-axis is almost one order of magnitude higher in the rotate-translate mode than in the translation-only mode. This results correspond to the increase of angular aperture associated with the rotation of the array (27).

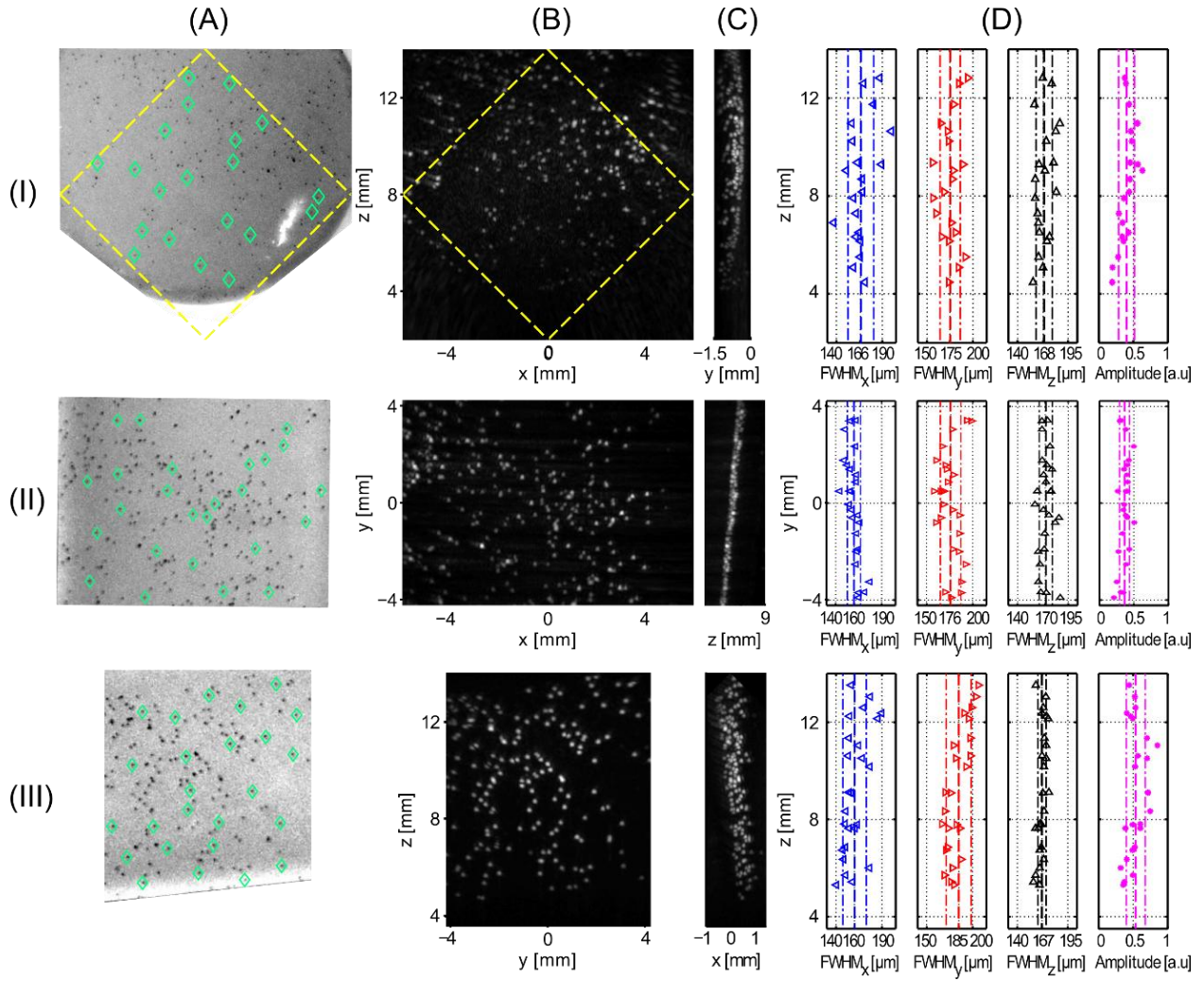

Fig. 3. Resolution of isotropic objects (Ø 50μm microspheres randomly spread on planar surfaces) for acquisitions in the rotate-translate mode. The rows (I)-(III) correspond to Phantom 1, Phantom 2 and Phantom 3, respectively. The phantoms are oriented so that the absorbing spheres are mainly comprised in planes perpendicular to the main axes of the coordinate system. Column (A): pictures of the phantoms, taken before embedding thin layer in a larger turbid gel. The green diamond markers indicate isolated (well-separated) microspheres that were chosen to assess the resolution. Column (B): MAP images along the y-axis (I), the z-axis (II) and the x-axis (III) of the 3D PA reconstructions, respectively. These MAP images correspond to the pictures in Column (A), and have the same scale. The yellow diamond on row (I) indicates the area where the angular aperture resulting from the motion of the array is expected to be maximal. Column (C): MAP images along the x-axis (I), the x-axis (II) and the y-axis (III) of the 3D PA reconstructions, respectively. These MAP images show that the microspheres are mainly constrained to a plane, and give its coordinate. Column (D): Results of the 3D fitting of each marked microsphere with a 3D Gaussian blob. The FWHM along the x, y and z-axis as well as the amplitude of the Gaussian are displayed. The amplitudes are normalized by the maximum value in the 3D images. For each plot, the dashed line indicates the mean value and the dashed-dotted lines the mean value ± the standard deviation over the 20 (I) or 25 (II)-(III) values. For all the FWHM plots the abscissa axis covers a 75 µm range.

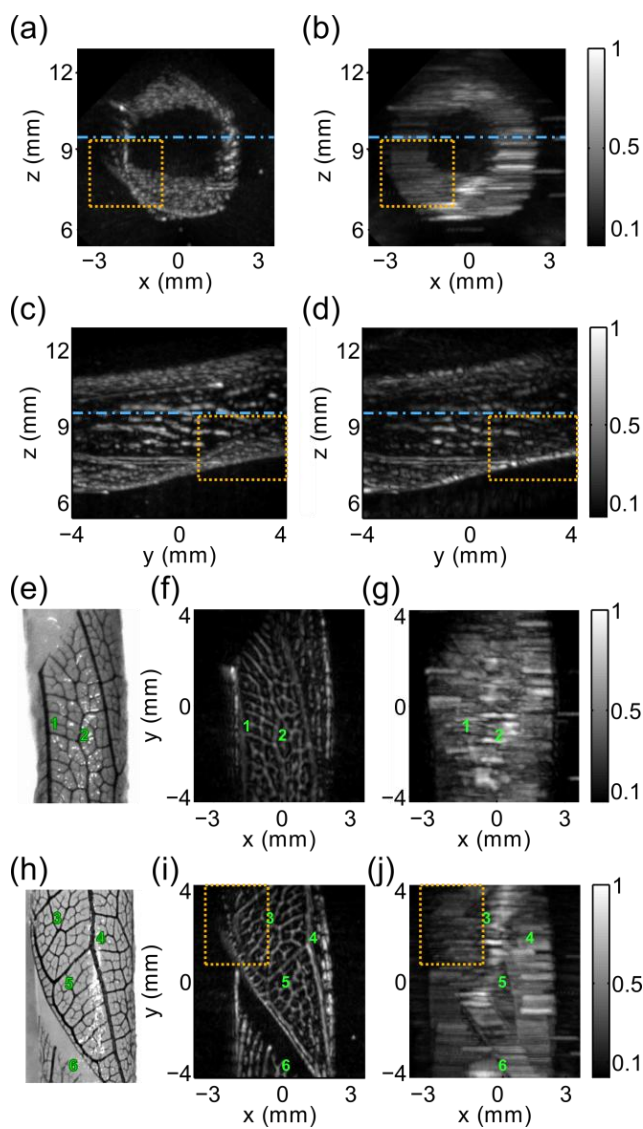

*3.2. Structures with multiple orientations: vascular network of a leaf skeleton*

Fig. 4 Reconstruction of a complex-shaped 3D object: a leaf skeleton. Maximum amplitude projection (MAP) images of Phantom 4, reconstructed with datasets acquired in the single-wavelength rotate-translate mode (a), (c), (f), (i) and the translate-only mode (b), (d), (g), (j), are presented (Video 1). (a) and (b) MAP images along the y-axis. (c) and (d) MAP images along the x-axis (Video 1). The dashed-dotted blue line marks the partition between two sub-volumes. The partition was performed to improve the visibility for the MAP images along the z-axis. (e) a photograph of the absorbing features for z >9.5mm and (f)-(g) the corresponding PA MAP images along the z-axis. (h) a photograph of the absorbing features for z <9.5mm and (i)-(j) the corresponding PA MAP images along the z-axis. The green numbers on (e)-(j) indicate positions of vessels nodes for easier comparison between the images. Each image was normalized by its maximum. The dotted orange lines mark the contour of a subset cuboid volume. (Video 1, QuickTime, 2.5 MB)

Fig. 4 and Video 1 present PA images of Phantom 4, composed of a strip of black leaf skeleton arranged in a helical ribbon shape, for acquisitions in the single-wavelength translate-only mode and in the single-wavelength rotate-translate scanning mode. Optical pictures of portion of the sample (Fig. 4 (e),(h)), taken before embedding the sample, show that the branching vessels of the leaf skeleton do not have any preferred in-plane orientation and the helical ribbon shape enable to explore multiple orientations in 3D.

The comparison between the images obtained in the two acquisition modes first shows that, as mentioned in section 3.1, the resolution along the x-axis is much higher in the rotate-translate scanning mode than in the translation-only one. This phenomenon is visible in particular in Fig. 4 (a) and (b) on the portion of the image around x = 2 mm. The low resolution in the translation-only acquisition mode makes it quite impossible to distinguish between the smeared vessels in the MAP images along the z-axis (Fig. 4 (g) and (j)). On the contrary, MAP images along the z-axis corresponding to the rotate-translate mode (Fig. 4 (f) and (i)) show that vessels of different sizes and orientation can be distinguished and their arrangement can be validated with the optical pictures. The reconstruction demonstrated that the developed PA rotate-translate system was able to resolve at least five orders of vessel branching.

MAP images along the x-axis are more similar between the two modes, especially in the central part of the images corresponding to portions of the sample almost parallel to the yz-plane (Fig. 4 (c) and (d)). The limited view linked to the finite length of the array and the presence of directive objects results in the partial visibility of the vessels for both modes. Vessels mostly parallel to the y-axis are visible. The rotation has only a small impact in the visibility of vessels for portions of the sample almost parallel to the yz-plane. However, a significant difference of visibility can be noticed between the MAP images along the x-axis in the dotted rectangle. This part of the images corresponds to a portion of the helical ribbon shape with a tangent plane in between the yz-plane and the x-plane (Fig. 4 (a), (h) and (i)). The rotate-translate scanning mode enabled reconstructing vessels with multiple orientations for this portion of the sample, because of the synthetic increase of the detection aperture. The small angular aperture for the ultrasound array did not allow reconstruction of directive structures with such orientations in the translation-only scanning mode. The effect of the synthetically increased detection aperture in the rotate-translate acquisition mode can also be seen in Video 1 displaying rotating MAP images around the y-axis with 2° angle between the projections.

*3.3.Multispectral*

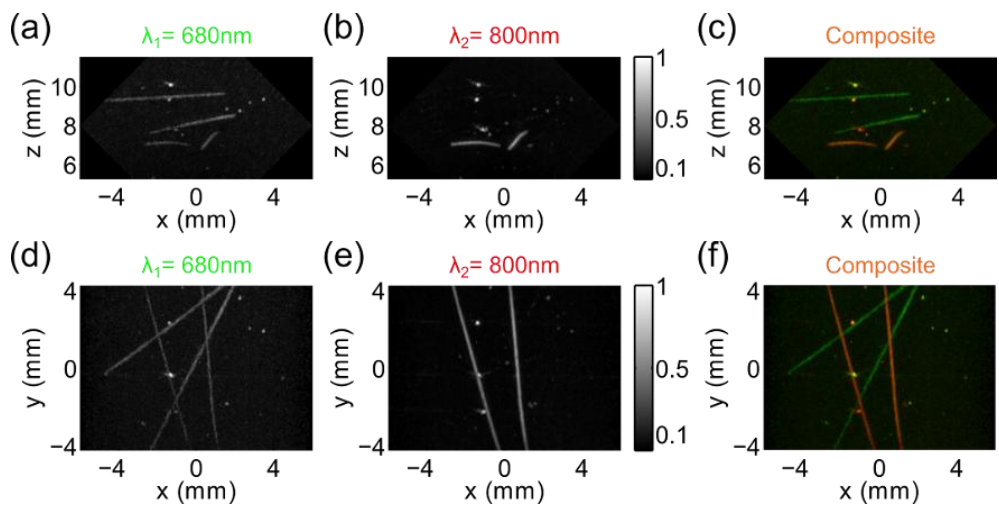

Fig. 5 Multispectral reconstruction of 3D objects: tubes filled with blue and black dyes (Video 2). Concentrations were adjusted so that the solution have the same optical density (OD=5) at $\lambda_1$=680nm. At $\lambda_2$=800nm the absorption of the blue ink is negligible compare to the absorption of the black ink. (a) and (b) MAP images along the y axis for pulses at $\lambda_1$ and $\lambda_2$ respectively. (d) and (e) MAP images along the z-axis for pulses at $\lambda_1$ and $\lambda_2$ respectively. Each image was normalized by its maximum. (c) and (f) color composite images of (a) and (b) or (d) and (e), respectively. (a) and (d) are displayed in green color scale and (b) and (e) in red color scale. (Video 2, QuickTime, 2.5 MB)

Phantom 5 comprising two capillary tubes filled with black indian ink and two capillary tubes filled with brilliant blue was imaged in the dual-wavelength rotate-translate scanning mode to demonstrate the multispectral capabilities of the PA imaging system. Two 3D PA images, each corresponding to laser shots at a different wavelength, were reconstructed (Fig. 5 and Video 2). At 680 nm, the four tubes are visible as the optical density (OD) of the black ink and the blue dye solution are similar. At 800 nm, however, only the two tubes filled with black ink can be observed clearly. This result was expected since the OD of the blue dye solution is 32 times lower than the OD of the black ink. Color composite images obtained by overlaying the two 3D images in different color scales (Fig. 5 (c) and (f), Video 2 right) show that the location of the tubes and the surrounding dust particles match between the two wavelengths. Moreover, the 3D image for each of wavelength has the same resolution and angular visibility. The difference in term of contrast can be attributed to the low power of the laser at 680 nm compare to 800 nm, which induce a lower signal-to-noise ratio. Fig. 5 and Video 2 demonstrate the ability of the system to discriminate in space between absorbers with two different spectra, which is an important starting point towards multispectral 3D imaging with chromophores having more complex spectra.

## 4. Discussion

We investigated in this study the performance of a novel rotate-translate scanner for multispectral volumetric PA imaging at mesoscopic scale. Emphasis was placed on operating with a high-frequency ultrasound array and electronics allowing simultaneous acquisition of all the detector elements of the array in a single-side access, planar detection geometry, but without being restricted to the angular aperture of the directional detectors. The study demonstrated that the proposed implementation provides a system able to image a cubic volume of side length 8.5 mm with a quasi-isotropic uniform 3D resolution of ~170 µm in a few minutes. Despite the limited angular aperture inherent to single side access PA imaging systems, rotation of the array was shown to both improve the resolution in the translation direction by almost one order of magnitude and extend the visibility of some directional structures, thereby increasing the image fidelity.

Regarding the field-of view achieved with the high-frequency implementation, a penetration depth of about one centimeter from the detection plane was reached. This depth corresponds to the depth-of-field of the 45° inclined array and is adapted to mesoscopic scale imaging. For the transverse directions, the image dimension along y-axis was limited by the length of the array; however, along x-axis the image could have been extended without degrading the image quality by using a longer translation range. Indeed, with the proposed scanning approach, the angular aperture in the translation direction is only dictated by the rotation range and therefore is independent of the translation range in the overlapping subset of all volumes imaged with the successive translation scans. This property makes our scanning approach more versatile than multi-directional scans based on a rotation of the array around the bisector in the median plane (33), which are limited by the angular aperture and length of the array in both of the transverse directions.

In a full-view detection geometry with omni-directional detectors, the expected resolution would have been on the order of the ultrasonic wavelength for the upper cutoff frequency limit of the ultrasonic detection (21), that is to say 75 µm (for 20MHz) here. Besides the limited angular aperture, several issues related to the reconstruction algorithm can explain the discrepancy between the measured resolution and the minimum detected ultrasonic wavelength. First, the electric impulse response (EIR) of the detection channel induced a spreading in time of the measured signals. The envelop detection method enabled to obtain images with positive values only, but transformed the time spread in a degraded resolution, in particular along the z-direction. This issue can be solved by deconvolving the measured signals with the EIR during the reconstruction process (44, 45), but requires challenging accurate measurement of the EIR at high frequency (46). Second, the directivity of the detector and the resulting non-uniform spatial impulse response (SIR) can induce low-pass filtering of the PA signals (46) and thereby degrade the resolution. Furthermore, assuming point-like detectors and ignoring their actual SIR was shown in other detection geometries to induce spatially-variant smearing and thereby to reduce the amplitude of some absorbing structures in the image (35, 47).While not taken into account in the simple radial back-projection reconstruction algorithm used here for image reconstruction, the SIR of the linear detector array can be included in advanced 3D reconstruction methods and was shown to yield superior imaging quality (35). In addition to improving the resolution, model-based reconstructions incorporating the SIR of the detectors have been demonstrated to reduce background noise of the images and in particular the radial streaks (visible in figure 1 (I-B)) caused by projections outside of the region of higher sensitivity of the detectors (35, 48). Model-based reconstructions are however computationally demanding due to the iterative process and the vast amount of data needed to generate high-resolution volumetric PA images. Moreover, no closed-form analytical expression can be found for the SIR of cylindrically focused detector, which results in additional computation complexity (35, 49). Practical implementations of more advanced reconstruction methods accounting for the EIR and the SIR of the detectors to improve image resolution are beyond the scope of this paper, but are of great interest for our rotate-translate system and will be considered in future studies.

Dual-wavelength imaging was demonstrated with ultrafast wavelength shifting. Interleaving acquisitions at different wavelengths was performed so that, in case of a slow motion of the sample, a similar distortion would affect the 3D images reconstructed for the different wavelengths. High resolution imaging implies high sensitivity to small motions or deformation of the sample. For spectroscopic PA imaging, a potential spatial mismatch between images acquired sequentially for the different excitation wavelengths can result in voxels exhibiting misleading absorption spectra (38). Although a dual-wavelength acquisition was adapted to discriminate between two absorbers with very different absorption spectra, our approach can easily be extended to a large number of wavelengths. Multi-wavelength un-mixing algorithms have been developed for discrimination between absorbers (50), and can even be used to perform denoising of the PA images (51). Imaging the complex distributions of multiple contrast agents with our system is beyond the scope of this paper, but will be investigated in the near future. Indeed, multispectral 3D PA imaging with a single-side access and an ultrafast wavelength shifting laser has been recently demonstrated to be of great interest with a imaging system operating at ultrasound frequencies up to 5MHz (52).

The spatial and temporal imaging performance of our PA scanner would be well suited for longitudinal studies in small animal models for human pathology, and for clinical application in dermatology or monitoring of superficial tumors. In the present paper, well-controlled imaging phantoms were used to demonstrate the validity of our approach *in vitro*. Considering this first implementation, several technical adjustments could be performed to facilitate future *in vivo* investigations. In particular, data-recording and positioning of the sample may be upgraded for easier acquisitions. Controllers with extended internal memory would avoid segmenting the scan in several parts, and respiration gating can be considered to eliminate excessive motion artifacts due to breathing (53). Appropriate sample holders should also be developed to immerse the surface of the imaged region ensuring a good acoustic coupling without the need for detector-to-body contact, and to prevent tremor and larger motions. For the illumination of samples accessible from one side only, two configurations could be envisioned. First, the illumination could be fixed, placed on the lateral sides (along the y-axis) of the moving array and directed towards the region of interest. Second, the optical fiber bundle could be attached to the ultrasound array to illuminate preferentially the median imaging plane as in (13, 14, 24, 33). However, for this second configuration, a reconstruction algorithm accounting of the varying illumination should be used (54). In biological tissue, acoustic attenuation of high ultrasound frequencies is expected to reduce the image quality with the penetration depth (55), but could be partially compensated using time-variant filtering (56).

The use of a conventional linear array and an ultrafast programmable ultrasound machine in our system offers the possibility to perform additional anatomical and functional imaging in the same configuration. Co-registration of PA imaging with ultrasonic imaging has been shown to be of great interest as it results in a more complete tissue characterization (57). Moreover, ultrafast ultrasound imaging (58) enables high frame rate imaging and can be used to acquire high quality ultrasound images between successive laser pulses. Our rotate-translate scanner could then be used to obtain high resolution 3D ultrasound images. Development of appropriate reconstruction algorithms will be considered in the frame of previously proposed reconstruction schemes (59). Furthermore, ultrafast ultrasound imaging has been shown to lead to superior Doppler imaging enabling tomographic reconstruction of the 3D vascular network based on vascular flow (60). For *in vivo* imaging, the combination of PA, ultrasound and ultrafast Doppler 3D imaging acquired in a single rotate-translate scan could give access to multiple biomarkers at the same time.

As photoacoustic imaging and multi-channel programmable ultrasound systems compatible with ultrasound linear arrays of various high center frequencies enter the market for biomedical imaging, our approach is expected to have an impact in both preclinical and clinical studies. With multispectral capabilities and acquisition time of a few minutes, single-side access rotate-translate scanners are ideal for repetitive fast measurements requiring 3D high resolution, and would certainly benefit longitudinal studies and therapy monitoring.

## 5. Acknowledgment

This work was funded by the Plan Cancer 2009–2013 (Action 1.1, Gold Fever), the Agence Nationale de la Recherche (Golden Eye, ANR-10-INTB-1003), the French program "Investissement d'Avenir" run by the Agence Nationale pour la Recherche (Infrastructure d'avenir en Biologie Santé, ANR-11-INBS-0006), and the LABEX WIFI (Laboratory of Excellence ANR-10-LABX-24) within the French Program "Investments for the Future" under reference ANR-10- IDEX-0001-02 PSL*. The authors thank Abdelhak Souilah for his support in the mechanical design, and Alexandre Perraud for his work on motor programming and assessment of the detector positions at the early stage of the system development, and on the mechanical design.

**This manuscript has been published as : J Gateau, M Gesnik, JM Chassot, E Bossy; "Single-side access, isotropic resolution, and multispectral three-dimensional photoacoustic imaging with rotate-translate scanning of ultrasonic detector array," J. Biomed. Opt., 20(5), 056004 (2015). The journal version is the definitive version.**34. J. Gateau, A. Chekkoury and V. Ntziachristos, "High-resolution optoacoustic mesoscopy with a 24 MHz multidetector translate-rotate scanner," *Journal of Biomedical Optics* **18**(10), 106005-106005 (2013)
35. M. A. Araque Caballero, J. Gateau, X. L. Dean-Ben and V. Ntziachristos, "Model-Based Optoacoustic Image Reconstruction of Large Three-Dimensional Tomographic Datasets Acquired With an Array of Directional Detectors," *Medical Imaging, IEEE Transactions on* **33**(2), 433-443 (2014)
36. S. Preisser, N. L. Bush, A. G. Gertsch-Grover, S. Peeters, A. E. Bailey, J. C. Bamber, M. Frenz and M. Jaeger, "Vessel orientation-dependent sensitivity of optoacoustic imaging using a linear array transducer," *Journal of Biomedical Optics* **18**(2), 026011-026011 (2013)
37. R. Ma, A. Taruttis, V. Ntziachristos and D. Razansky, "Multispectral optoacoustic tomography (MSOT) scanner for whole-body small animal imaging," *Opt. Express* **17**(24), 21414-21426 (2009)
38. X. L. Deán-Ben, E. Bay and D. Razansky, "Functional optoacoustic imaging of moving objects using microsecond-delay acquisition of multispectral three-dimensional tomographic data," *Sci. Rep.* **4**((2014)
39. A. Buehler, M. Kacprowicz, A. Taruttis and V. Ntziachristos, "Real-time handheld multispectral optoacoustic imaging," *Optics letters* **38**(9), 1404-1406 (2013)
40. A. N. S. I. American National Standard for Safe Use of Lasers, Z136.1, 2000,
41. M. Xu and L. V. Wang, "Universal back-projection algorithm for photoacoustic computed tomography," *Physical Review E* **71**(1), 016706 (2005)
42. C. A. Schneider, W. S. Rasband and K. W. Eliceiri, "NIH Image to ImageJ: 25 years of image analysis," *Nat Meth* **9**(7), 671-675 (2012)
43. M. I. Khan and G. J. Diebold, "The photoacoustic effect generated by an isotropic solid sphere," *Ultrasonics* **33**(4), 265-269 (1995)
44. W. Yi, X. Da, Z. Yaguang and C. Qun, "Photoacoustic imaging with deconvolution algorithm," *Physics in medicine and biology* **49**(14), 3117 (2004)
45. A. Rosenthal, V. Ntziachristos and D. Razansky, "Acoustic Inversion in Optoacoustic Tomography: A Review," *Current medical imaging reviews* **9**(4), 318 (2013)
46. M. A. A. Caballero, A. Rosenthal, A. Buehler, D. Razansky and V. Ntziachristos, "Optoacoustic determination of spatio- temporal responses of ultrasound sensors," *Ultrasonics, Ferroelectrics, and Frequency Control, IEEE Transactions on* **60**(6), 1234-1244 (2013)
47. K. Mitsuhashi, K. Wang and M. A. Anastasio, "Investigation of the far-field approximation for modeling a transducer's spatial impulse response in photoacoustic computed tomography," *Photoacoustics* **2**(1), 21-32 (2014)
48. K. Wang, R. Su, A. A. Oraevsky and M. A. Anastasio, "Investigation of iterative image reconstruction in three-dimensional optoacoustic tomography," *Physics in medicine and biology* **57**(17), 5399-5423 (2012)
49. D. Queirós, X. L. Déan-Ben, A. Buehler, D. Razansky, A. Rosenthal and V. Ntziachristos, "Modeling the shape of cylindrically focused transducers in three-dimensional optoacoustic tomography," *Journal of Biomedical Optics* **18**(7), 076014-076014 (2013)
42